\documentclass[a4paper,fleqn,usenatbib]{mnras}
\usepackage{graphicx}

\newcommand{\zab}{\ensuremath{z_\textrm{\scriptsize abs}}}
\newcommand{\zem}{\ensuremath{z_\textrm{\scriptsize{em}}}}
\newcommand{\dmm}{\Delta\mu/\mu}
\newcommand{\kms}{km\,s$^{-1}$}
\newcommand{\ms}{m\,s$^{-1}$}

\newcommand\adp{Ann.\ Phys.}

\newcommand\jms{J. Mol.\ Spectrosc.}
\newcommand\jpb{J. Phys.\ B}

\newcommand\lrr{Living Rev. Relat.}

\newcommand\soviet{Sov. J. Exp. Theor. Phys.}
\newcommand\ieee{IEEE Trans. Aut. Contr.}
\newcommand\molphys{Mol. Phys.}
\newcommand\canjphys{Can.\ J.\ Phys.}
\newcommand\revmodphys{Rev. Mod. Phys.}
\newcommand\jqsrtrans{J.\ Quant.\ Spectrosc.\ Radiat.\ Transfer}

\title[Constraint on a varying $\mu$ at $\zab \simeq 2.34$]{Constraint on a varying proton-to-electron mass ratio from H$_{2}$ and HD absorption at $\zab \simeq 2.34$}

\author[Dapr\`a et al.]{M. Dapr\`a,$^{1}$ M. van der Laan,$^{1}$ M. T. Murphy,$^{2}$ and W. Ubachs,$^{1}$\\
$^{1}$Department of Physics and Astronomy, LaserLaB, VU University,
  De Boelelaan 1081, 1081 HV Amsterdam, The Netherlands\\
$^{2}$Centre for Astrophysics and Supercomputing, Swinburne University of Technology, Melbourne, Victoria 3122, Australia\\
}

\begin{document}

\date{}

\pagerange{\pageref{firstpage}--\pageref{lastpage}} \pubyear{2015}

\maketitle

\label{firstpage}

\begin{abstract}
Molecular hydrogen absorption in the damped Lyman-$\alpha$ system at $\zab = 2.34$ towards quasar Q\,1232+082 is analyzed in order to derive a constraint on a possible temporal variation of the proton-to-electron mass ratio, $\mu$, over cosmological timescales. Some 106 H$_{2}$ and HD transitions, covering the range \mbox{3290-3726 \AA}, are analyzed with a comprehensive fitting technique, allowing for the inclusion of overlapping lines associated with hydrogen molecules, the atomic hydrogen lines in the Lyman-$\alpha$ forest as well as metal lines. The absorption model, based on the most recent and accurate rest wavelength for H$_{2}$ and HD transitions, delivers a value of \mbox{$\dmm = (19 \pm 9_{stat} \pm 5_{syst}) \times 10^{-6}$}. An attempt to correct the spectrum for possible long-range wavelength distortions is made and the uncertainty on the distortion correction is included in the total systematic uncertainty. The present result is an order of magnitude more stringent than a previous measurement from the analysis of this absorption system, based on a line-by-line comparison of only 12 prominent and isolated H$_{2}$ absorption lines. This is consistent with other measurements of $\dmm$ from 11 other absorption systems in showing a null variation of the proton-to-electron mass ratio over a look-back time of \mbox{$11$ Gyrs}.
\end{abstract}


\section{Introduction}
\label{sec:intro}
The search for possible variations in the values of fundamental constants over the course of cosmic history from spectroscopic observations of highly redshifted galaxies has become an established field of research in the past decade. In particular, drifts in dimensionless constants, such as the  fine-structure constant $\alpha = e^{2}/(4 \pi \epsilon_{0} \hbar c)$ and the proton-to-electron mass ratio $\mu = M_{p}/m_{e}$, can be probed by observing and calibrating atomic and molecular absorption lines from objects in the early Universe, and comparing their wavelengths with those measured in the laboratory~\citep{Uzan2011,Ubachs2016}. 

Temporal drifts of the proton-to-electron mass ratio, a dimensionless constant responsible of the structure of molecular matter, can be probed using transitions of a variety of molecules \citep{Kozlov2013b,Jansen2014}. Astronomical observations in the radio domain targeting ammonia \citep{Murphy2008,Kanekar2011} and methanol \citep{Bagdonaite2013a,Bagdonaite2013b,Kanekar2015} delivered tight constraints on $|\dmm|$ at the level of $10^{-7}$. However, ammonia is detected in only two extragalactic systems, B0218+357 \citep{Henkel2005} at \mbox{$\zab \simeq 0.7$} and PKS1830-211 \citep{Henkel2008} at \mbox{$\zab \simeq 0.9$}, while methanol is detected only in the latter \citep{Muller2011}.

\cite{Thompson1975} first proposed to use molecular absorption in high redshift systems to probe the variation of $\mu$. Molecular hydrogen, H$_{2}$, is found to have many transitions showing different sensitivities within its band systems \citep{Varshalovich1993,Ubachs2007} and it is a good candidate to investigate a possible $\mu$-variation. Being the most abundant molecule in the Universe, H$_{2}$ is found in a larger number of absorption systems, which display up to 100 H$_{2}$ absorption features. The Lyman and Werner band systems can be observed using the large, ground-based telescopes at $\zab > 2$, for which they  are redshifted into the atmospheric transmission window \mbox{($\lambda > 3050$ \AA)}. Nine \mbox{H$_{2}$-bearing} absorption systems in the range \mbox{$\zab = 2.05 - 4.22$} were analyzed recently for $\mu$-variation, delivering an average constraint of \mbox{$|\dmm| < 5 \times 10^{-6}$} at $3\sigma$ level \citep[see][and references therein for the single absorbers]{Ubachs2016}.

The presence of molecular hydrogen in the damped Lyman-$\alpha$ (DLA) absorption system at $\zab \simeq 2.34$ in the line of sight towards quasar SDSS J123437.55+075843.3, hereafter Q1232+082, was first reported by \cite{Ge1999} from medium-resolution observations. Based on observations of Q1232+082 with the Ultraviolet and Visual Echelle Spectrograph (UVES) mounted on the \mbox{8.2 m} Very Large Telescope (VLT), \cite{Srianand2000} deduced a value for the cosmic microwave background temperature at the local redshift from the fine-structure lines of \mbox{C \textsc{i}} combined with H$_{2}$ lines. A measurement of a varying $\mu$ based on the analysis of 12 unsaturated and isolated H$_{2}$ absorption features was reported by \cite{Ivanchik2002}. They derived two different values constraining $\dmm$ at the level of \mbox{$|\dmm| < 2 \times 10^{-4}$} ($3 \sigma$) using the H$_{2}$ laboratory wavelengths stemming from classical spectroscopic methods as reported by \cite{Morton1976} and \cite{Abgrall1993}. A later detailed analysis of the H$_{2}$ and HD absorption lines was reported by \cite{Ivanchik2010}, who found a ratio $N($HD$)/N($H$_{2})$ significantly larger than that found in the interstellar clouds in the Galaxy. Moreover, they reported that the absorbing cloud does not cover entirely the broad line region of the background quasar Q1232+082. Its partial coverage was further analyzed by \cite{Balashev2011}.

The main goal of this work is to perform an improved re-analysis of the H$_{2}$ absorption system Q1232+082 in order to constrain a temporal variation of the proton-to-electron mass ratio. To achieve this, the powerful comprehensive fitting technique \citep{King2008,Malec2010} was applied to the quasar spectrum. This technique allows the treatment of partially overlapped absorption features, increasing the number of transitions included in the sample. This, along with an improved set of rest wavelengths, considered exact for the purpose of the comparison, led to a lower statistical uncertainty on the derived $\dmm$ value. Finally, the absorption model includes H$_{2}$ and HD column densities that are corrected for the partial coverage of the absorbing cloud. The quasar observations used are presented in Section \ref{sec:data}, the fitting method is described in Section \ref{sec:method}, the results, including the measurement of $\dmm$, are presented in Section \ref{sec:analysis}, while the effect of systematics is discussed in Section \ref{sec:systematic}.

\section{Data}
\label{sec:data}
The exposures used in this work were taken in five different observational programs obtained from the ESO data archive\footnote{\url{http://archive.eso.org/eso/eso_archive_main.html}}. The programs, whose details are listed in \mbox{Table \ref{tab:obs}}, were carried out between 2000 and 2003 using UVES, for a total integration time on the target of \mbox{19.4 hrs}. During all the observations, the blue arm of UVES was centred on \mbox{390 nm} in order to cover the H$_{2}$ Lyman and Werner series window. Most of the exposures had a slit width of 1.0 arcsec and a CCD binning of \mbox{2$\times$2} (spectral $\times$ spatial), with the only exception of program \mbox{70.A-0017(A)}, which had a slit width of 1.2 arcsec and a binning of \mbox{3$\times$2}.

The raw 2D exposures were reduced following the same procedure described by \cite{King2011} and \cite{Bagdonaite2012,Bagdonaite2014}. The exposures were flat-fielded and bias-corrected using the Common Pipeline Language (CPL) version of the UVES pipeline. Subsequently, the CPL was used to optimally extract the quasar flux. Since none of the programs was carried out with attached ThAr calibration exposures, the single quasar exposures were wavelength calibrated using the standard ThAr exposures taken at the end of the night. After the standard reduction, the echelle orders were combined onto a common vacuum-heliocentric wavelength grid, with a dispersion of \mbox{2.5 \kms} per pixel, in a single normalized spectrum using the custom software \textsc{UVES\_popler} \citep{UVESpopler}. The flux density and variance was linearly redispersed from the original pixel scale to the new wavelength grid in the same way for all exposures, including the single \mbox{3$\times$2}-binned exposures. The flux arrays for all orders were scaled to optimally match each other where they overlapped in wavelength space and the final spectrum was formed using an inverse-variance weighted mean of the contributing exposures at each pixel. The combined 1D spectrum was manually inspected and bad pixels and other spectral artifacts were removed. Subsequently, the quasar continuum was fitted with low-order polynomials. 

The final spectrum of Q1232+082 (resolving power $R \sim 45000$) covers the wavelengths from 3290 to \mbox{6650 \AA}, with gaps between \mbox{4525-4620 \AA} and \mbox{5598-5672 \AA} due to the separation between the CCDs. The signal-to-noise ratio (S/N) is $\sim 20$ per 2.5 \kms\ per pixel at \mbox{$\sim 3500$ \AA}, in the middle of the H$_{2}$ window in the quasar spectrum.
\begin{table*}
    \centering
    \caption{Observational details of the Q1232+082 exposures with UVES/VLT obtained from the ESO archive and used in this work. The grating settings were \mbox{$390+564$ nm} for each exposure.}
    \label{tab:obs}
    \begin{tabular}{cccccc}
    \hline
    Program ID & Date & Execution & Integration & Slit width & CCD     \\
               &      & time (UT)       & time [s]    & [arcsec]   & binning \\
    \hline
    \hline
    65.P-0038(A) & 06/04/2000 & 04:11:32 & 3600 & 1.0 & 2$\times$2 \\
                 & 06/04/2000 & 05:15:52 & 3600 &     &            \\
                 & 08/04/2000 & 05:26:42 & 3600 &     &            \\
    68.A-0106(A) & 08/01/2002 & 06:57:28 & 6000 &     & 2$\times$2 \\
                 & 09/01/2002 & 06:56:17 & 6000 &     &            \\
                 & 10/01/2002 & 06:59:36 & 6000 &     &            \\
    69.A-0061(A) & 02/06/2002 & 00:16:07 & 5400 &     & 2$\times$2 \\
                 & 10/06/2002 & 00:46:31 & 5400 &     &            \\
    70.A-0017(A) & 04/01/2003 & 07:06:19 & 5200 & 1.2 & 3$\times$2 \\
    71.B-0136(A) & 02/04/2003 & 02:55:36 & 5400 & 1.0 & 2$\times$2 \\
                 & 02/04/2003 & 04:29:52 & 5400 &     &            \\
                 & 03/04/2003 & 03:37:30 & 3600 &     &            \\
                 & 03/04/2003 & 04:39:58 & 3600 &     &            \\
                 & 02/05/2003 & 03:00:06 & 3600 &     &            \\
                 & 02/05/2003 & 04:02:07 & 3600 &     &            \\
    \hline
    \end{tabular}
\end{table*}

\section{Method}
\label{sec:method}
The fitting technique used in this work is the comprehensive fitting method introduced by \cite{King2008} and later refined by \cite{Malec2010}. This method involves a simultaneous treatment of all the considered transitions. The main strength of this technique is that the fitting parameters used to describe the absorption features can be tied together. This results in a lower number of free parameters and allows the H$_{2}$ lines that are partially overlapped by or blended with intervening spectral features to be modelled.

The absorption model was created using \textsc{vpfit} \citep{vpfit}, a custom nonlinear least-squares Voigt profile fitting program developed specifically for quasar absorption. The absorption feature profiles were fitted using a Voigt profile, which consists in the convolution of a Lorentzian profile reflecting the natural line profile, which is specific for each transition considered, a Gaussian profile describing the broadening due to the thermal and turbulent velocities of the absorbing cloud, and an instrumental profile assumed to be Gaussian. In \textsc{vpfit}, each absorption feature is described by a set of 3 free parameters: the column density \emph{N}, the redshift at which the absorption occurs \zab, and the Doppler line width \emph{b}. 
The atomic and molecular properties of each transition included in the absorption model, namely the laboratory wavelength $\lambda^{0}$, the oscillator strength \emph{f}, the damping parameter $\Gamma$, and the sensitivity coefficient \emph{K}, are included in \textsc{vpfit} as fixed values. The laboratory wavelengths for the H$_{2}$ transitions were measured with fractional accuracies of \mbox{$\sim 5 \times 10^{-9}$} and \mbox{$1-2 \times 10^{-8}$} for Lyman and Werner transitions respectively by \cite{Salumbides2008}, while HD transitions were measured by \cite{Hollenstein2006} and \cite{Ivanov2008} with a relative accuracy of \mbox{$\sim 5 \times 10^{-8}$}. The oscillator strengths and the damping parameters were calculated by \cite{Abgrall1994}  and by \cite{Abgrall2000}, respectively, for H$_{2}$ transitions and by \cite{Abgrall2006} for HD transitions. The sensitivity coefficients were calculated via a semi-empirical analysis by \cite{Ubachs2007} for H$_{2}$, while they were derived in ab initio calculations by \cite{Ivanov2010} for HD transitions. The database used in this work was tabulated by \cite{Malec2010}.

The underlying assumption of this work is that absorption features detected at the same $\zab$ originate from the same absorbing cloud. In other words, they share the same physical conditions, like the turbulent motions and the temperature of the cloud. This is represented in \textsc{vpfit} by tying the redshift and the Doppler width parameters among the different transitions. Moreover, it is assumed that transitions probing the same rotational states are sharing the same level population. As a consequence, their column densities $N_{J}$ were tied together.

\textsc{vpfit} works iteratively, starting from user-supplied initial values for each free parameter in the absorption model. During each iteration, the values of such parameters are changed in order to minimize the value of the control parameter $\chi^{2}$. The program stops iterating once a stopping criterion, which is user-defined, is met, reporting convergence. Beside the chi-squared parameter, the Akaike Information Criterion \citep[AICC][]{akaike1974} is used to control the goodness of the fit. In particular, this parameter is defined as:
\begin{equation}
    AICC = \chi^{2} + 2p + \frac{2p(p + 1)}{n - p - 1},
    \label{eq:aicc}
\end{equation}
where \emph{p} is the number of free parameters, and \emph{n} is the number of the data points included in the fit. In particular, a difference of $\Delta AICC > 5$ between two absorption models is considered to be strong evidence that the model with the lower \emph{AICC} value is statistically preferred.

\section{Analysis}
\label{sec:analysis}
The spectrum of quasar Q1232+082 ($\zem = 2.57$) contains a DLA that features molecular hydrogen absorption, at $\zab = 2.33771$. The molecular hydrogen absorption in the system was investigated by \cite{Srianand2000} and \cite{Varshalovich2001}, who reported the first deuterated molecular hydrogen (HD) detection in a DLA. A later, combined study of H$_{2}$ and HD absorption found evidence of partial coverage in the absorbing system \citep{Ivanchik2010}.

The presence of partial coverage means that the angular radius of the absorbing cloud is smaller than that of the background source. As a consequence, even saturated absorption features originating in the cloud do not absorb all the quasar radiation, leaving a residual flux in the spectrum. The partial coverage is characterized by a covering factor $f = F_{abs}/F_{tot}$, which is defined as the ratio between the flux affected by the cloud absorption, $F_{abs}$, and the total flux of the background source, $F_{tot}$. If all the quasar radiation is covered by the cloud, then the covering factor is unity. This results in saturated lines going to the zero level in the spectrum. A covering factor $f < 1$ means that there is a residual flux from the background source which prevents saturated lines to go to the zero level, as shown in \mbox{Fig. 1} of \cite{Ivanchik2010}. It is worth noting that different elements/molecules' absorption features originate in different parts of the absorbing cloud, hence their covering factors are in principle different. The partial coverage phenomenon can be explained by invoking a number of different effects \citep[e.g.][]{Balashev2011}, for example an effectively smaller radius of the absorbing cloud or the presence of multiple, unresolved background sources illuminating the absorber. However, to explain the causes of the partial coverage in this system is beyond the goal of this work. A detailed study of the partial coverage of the Q1232+082 broad line region, considering multiple elements, was reported by \cite{Balashev2011}.

Molecular hydrogen is detected in the spectrum in its two forms, H$_{2}$ and HD, with more than 100 lines detected for observed wavelengths shorter than \mbox{3726 \AA}. The dataset considered in this analysis was built considering only the transitions probing the rotational states with $J \le 5$, and is presented in \mbox{Table \ref{tab:transitions}}. The $J = 0-1$ transitions have a large column density and are heavily saturated. Therefore, they were included in the analysis only when no evidence of overlap with strong, intervening \mbox{H \textsc{i}} lines was found. Since the H$_{2}$ and HD transitions fall in the Lyman-$\alpha$ forest, overlaps with intervening neutral hydrogen and/or metal transitions are common. Within the comprehensive fitting method, partial overlaps can be handled, therefore such transitions were included in the dataset. However, molecular hydrogen transitions that are completely overlapped by saturated \mbox{H \textsc{i}} lines were excluded from the dataset, since they do not add relevant information to the signal. 
In total, 96 H$_{2}$ transitions, belonging to the Lyman and the Werner band systems, and 10 HD transitions were selected among 42 spectral regions in the range \mbox{3290-3726 \AA}.
\begin{table*}
    \centering
    \caption{List of the 106 molecular hydrogen transitions considered in this work.}
    \label{tab:transitions}
    \begin{tabular}{cp{8cm}p{5cm}c}
    \hline
    \emph{J}-level & Lyman transitions & Werner transitions & \# \\
    \hline
    \hline
        0 & L0R(0), L1R(0), L2R(0), L4R(0),  &  & 4 \\
        1 & L0P(1), L0R(1), L1P(1), L1R(1), L2P(1), L2R(1), L3P(1), L4P(1), L4R(1), L5P(1), L7P(1) &  & 11 \\
        2 & L0P(2), L0R(2), L1P(2), L1R(2), L2P(2), L2R(2), L3P(2), L3R(2), L4P(2), L4R(2), L5P(2), L5R(2), L6P(2), L7P(2), L7R(2), L8P(2), L9R(2), L9P(2) & W0P(2), W1P(2), W1Q(2) & 21 \\
        3 & L0P(3), L0R(3), L1P(3), L1R(3), L2P(3), L2R(3), L3P(3), L3R(3), L4P(3), L4R(3), L5P(3), L5R(3), L6P(3), L6R(3), L7P(3), L7R(3), L8R(3), L9P(3), L9R(3), L10P(3), & W0P(3), W1Q(3), W1R(3) & 23 \\
        4 & L1P(4), L1R(4), L2P(4), L2R(4), L3P(4), L3R(4), L4P(4), L4R(4), L5P(4), L5R(4), L6P(4), L6R(4), L8P(4), L9R(4), L10R(4)  & W0P(4), W0Q(4), W0R(4), W1P(4)  & 20 \\
        5 & L1P(5), L1R(5), L2P(5), L2R(5), L3P(5), L3R(5), L5P(5), L5R(5), L6P(5), L7P(5), L8P(5), L9P(5), L10P(5), L10R(5) & W0R(5), W1Q(5)  & 17 \\
    \hline
        HD $J = 0$ & L1R(0), L2R(0), L3R(0), L4R(0), L5R(0), L6R(0), L7R(0), L8R(0) & W0R(0), W1R(0) & 10 \\
    \hline
    \end{tabular}
\end{table*}

Each absorption feature was modelled by assigning it a set of free parameters in \textsc{vpfit}. The absorption redshift $\zab$ and the width \emph{b} were tied among all the H$_{2}$ transitions, while the column densities $N_{J}$ were tied only among the transitions probing the same rotational state. All the HD absorption features were described by a different set of free parameters, with only the redshift tied to the H$_{2}$ value. In order to build a robust absorption model, the Lyman-$\alpha$ forest \mbox{H \textsc{i}} lines falling in the selected spectral regions were modelled by assigning to each of them a set of free parameters. Note that the parameters describing these lines were not tied to any value and were allowed to vary independently from each other. The initial value of each fitting parameter was user-provided.

To account for possible misplacements of the global quasar continuum, a continuum correction was introduced in each spectral region. This correction acts as a local continuum, limited to the spectral region considered. In \textsc{vpfit} the local continua are fitted with a straight line to the spectrum, therefore diminishing the impact of a misplaced global quasar continuum on the absorption model. To account for the partial coverage, a zero-level correction was introduced in each spectral region. Working in a similar way than the continuum correction, this correction locally shifts the zero-level of the quasar spectrum. A covering factor of $f = 0.92 \pm 0.01$ was derived from the weighted average of the zero-level corrections of all the spectral regions considered. This value matches with that found by \cite{Ivanchik2010} for the H$_{2}$ lines.

\begin{figure}
    \centering
    \includegraphics[width=\columnwidth]{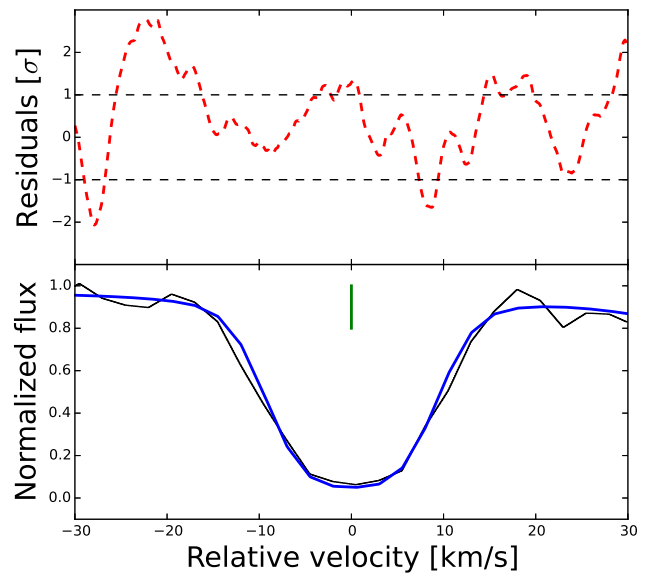}
    \caption{Top panel: normalized composite residual spectrum from 32 unblended and non-overlapping H$_{2}$ transitions. The dashed lines represent the $\pm1 \sigma$ boundaries. Bottom panel: the \mbox{H$_{2}$ L3R(3)} transition, with a rest wavelength of \mbox{$\lambda^{0} = 1067.5$ \AA}, is plotted as reference. The velocity scale is centred at the absorption redshift $z = 2.33771$. The solid (blue) line shows the absorption model, while the solid (green) tick shows the position of the VC.}
    \label{fig:crs_h2}
\end{figure}

\begin{figure}
    \centering
    \includegraphics[width=\columnwidth]{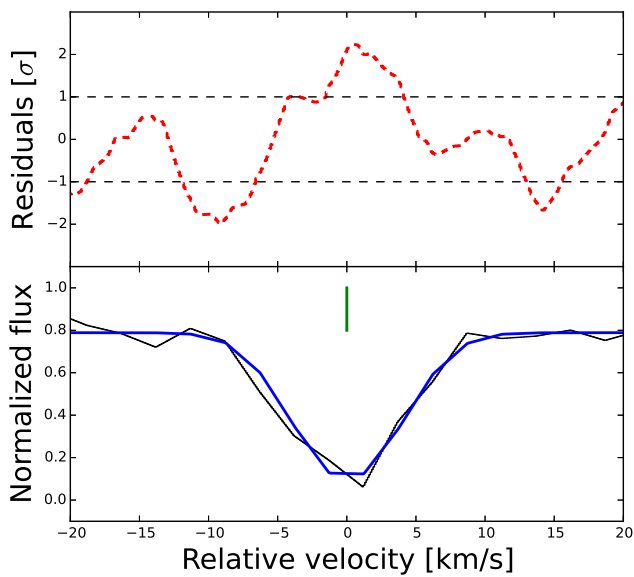}
    \caption{Top panel: normalized composite residual spectrum from 6 unblended and non-overlapping HD transitions. The dashed lines represent the $\pm1 \sigma$ boundaries. Bottom panel: the \mbox{HD 7R(0)} transition, with a rest wavelength of \mbox{$\lambda^{0} = 1021.5$ \AA}, is plotted as reference. The velocity scale is centred at the absorption redshift $z = 2.33771$. The solid (blue) line shows the absorption model, while the solid (green) tick shows the position of the VC.}
    \label{fig:crs_hd}
\end{figure}

Even though molecular hydrogen absorption features were fitted using a single Voigt profile, it is common to find more complex velocity structures underlying the H$_{2}$ absorption features \citep[e.g.][]{Malec2010,Bagdonaite2014,Dapra2015}. Therefore the presence of extra velocity components (VCs) was investigated. A composite residual spectrum \citep[CRS,][]{Malec2010} was built using single, isolated absorption features for both H$_{2}$ and HD. Combining the residuals of multiple transitions, the CRS exploits the presence/absence of VCs in the absorption model. The CRS relative to H$_{2}$ absorption features, presented in \mbox{Fig. \ref{fig:crs_h2}}, suggests a possible extra VC at \mbox{$\sim 8.5$ \kms}, with respect to the absorption redshift $\zab = 2.33771$. To test the presence of an extra VC, multiple models were fitted, each including a second VC for each H$_{2}$ \emph{J}-level. During the fitting process the second VC's column density and Doppler width values reached some user-defined limits, that were set to $\log[N/\mathrm{cm}^{-2}]<8.0$ and \mbox{$b < 0.05$ \kms}, hence it was rejected by \textsc{vpfit}. As a consequence, the single VC model was adopted. The same analysis was performed looking for extra VCs for the HD absorption features, but the CRS for HD, which is presented in \mbox{Fig. \ref{fig:crs_hd}}, does not show evidence of extra VCs.

The parameters of the single VC model are presented in \mbox{Table \ref{tab:fit_results}}, while the complete absorption model is compared to the observed spectrum in \mbox{Appendix \ref{app:model}}. The values of the H$_{2}$ column densities are in good agreement with those reported by \cite{Ivanchik2010}, with only the rotational states with $J = 2$ and 3 having a larger $N_{J}$ value, while the line width \emph{b} presented here is smaller. The difference between the two \emph{b} values may be explained by the fact that \cite{Ivanchik2010} derived their value from the curve of growth of rotational states with $J = 2-5$, while the value presented in this work is derived considering the contribution from rotational states with $J < 5$. The Doppler width parameter of the H$_{2}$ lines was observed to vary with the \emph{J}-level \citep[e.g. in HE0027--1836,][]{Noterdaeme2007}. According to \cite{Balashev2009}, this can imply that the H$_{2}$ transitions with low \emph{J} originate in the central part of the absorbing cloud, where the turbulence is minimal. On the other hand, high \emph{J} transitions originate in the external part of the cloud, where larger turbulence may contribute more to the line broadening. As a consequence, the Doppler width derived only from the high-\emph{J} states by \cite{Ivanchik2010} is larger than that reported here, which includes the contribution from the states with $J = 0$ and 1. In \mbox{Section \ref{subsubsec:untied}} a test is performed to investigate the impact of the assumption of spatial homogeneity on the $\dmm$ value derived in this work. The parameters describing the HD features are well in agreement with those reported by \cite{Ivanchik2010}.
\begin{table}
    \centering
    \caption{Column densities $N_{J}$ and Doppler width \emph{b} of the H$_{2}$ and HD transitions in the Q1232+082 spectrum.}
    \label{tab:fit_results}
    \begin{tabular}{ccc}
    \hline
    Rotational level & $\log[N_{J}$/cm$^{-2}]$ & \emph{b} [\kms]\\
    \hline
    $J = 0$ & $19.41 \pm 0.01$ & $3.39 \pm 0.06$ \\
    $J = 1$ & $19.26 \pm 0.01$ & \\
    $J = 2$ & $17.47 \pm 0.04$ & \\
    $J = 3$ & $17.16 \pm 0.05$ & \\
    $J = 4$ & $14.76 \pm 0.02$ & \\
    $J = 5$ & $14.23 \pm 0.02$ & \\
    \hline
    HD $J = 0$ & $15.74 \pm 0.53$ & $1.67 \pm 0.27$ \\
    \hline
    \end{tabular}
\end{table}

\subsection{Constraining $\dmm$}
\label{subsec:constraint}
\begin{figure}
    \centering
    \includegraphics[width=\columnwidth]{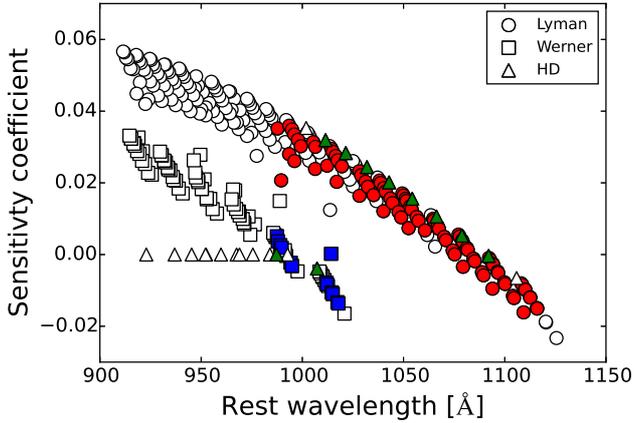}
    \caption{Sensitivity coefficients of Lyman and Werner transitions of H$_{2}$ and of HD transitions. The transitions considered in this analysis are marked with shaded markers.}
    \label{fig:ki}
\end{figure}

As proposed by \cite{Thompson1975}, the H$_{2}$ absorption detected in high-redshift systems can be used to detect a temporal variation of the proton-to-electron mass ratio $\mu$. A variation of $\mu$ is reflected by a shift in the observed wavelength $\lambda^{z}_{i}$ of the \emph{i}-th transition according to:
\begin{equation}
    \lambda^{z}_{i} = \lambda^{0}_{i} (1 + \zab) (1 + K_{i} \frac{\Delta \mu}{\mu}),
    \label{eq:shift}
\end{equation}
where $\lambda^{0}_{i}$ is the rest wavelength of the transition, $\zab$ is the absorption redshift, $\dmm \equiv (\mu_{z} - \mu_{0})/\mu_{0}$ is the relative difference between the proton-to-electron mass ratio in the absorption system and the value measured on Earth, and $K_{i}$ is  the sensitivity coefficient, defined as:
\begin{equation}
    K_{i} \equiv \frac{\mathrm d \ln \lambda^{0}_{i}}{\mathrm d \ln \mu}.
    \label{eq:ki}
\end{equation}
The coefficient $K_{i}$ determines the sign and magnitude of the sensitivity to a varying $\mu$ and is specific for the \mbox{\emph{i}-th} transition. The $K_{i}$ coefficients used in this work were calculated within a semi-empirical framework by \cite{Ubachs2007}, including the effects beyond the Born-Oppenheimer approximation and are presented in \mbox{Fig \ref{fig:ki}}.

The variation of $\mu$ is represented in \textsc{vpfit} by an extra free parameter. It was added to the other parameters ($N_{J}$, $\zab$, \emph{b}) of the H$_{2}$ and HD transitions only after having developed a robust absorption model. This fourth parameter was not added earlier in order to avoid that any flaw in the model itself would be compensated by an artificial variation of $\mu$. The model returned a value of the variation of the proton-to-electron mass ratio of \mbox{$\dmm = (22 \pm 9_{stat}) \times 10^{-6}$}. It is worth noting that the statistical error was derived from the diagonal term of the final covariance matrix, representing thereby only the uncertainty arising from the S/N of the spectrum.

\cite{Ivanchik2002} derived two values for $\dmm$ using the H$_{2}$ absorption in the system towards Q1232+082. They analyzed 12 isolated, unsaturated and unblended H$_{2}$ lines and used two independent sets of rest wavelengths $\lambda^{0}$ from \cite{Morton1976} and \cite{Abgrall1993} finding \mbox{$\dmm = (144 \pm 114) \times 10^{-6}$} and \mbox{$\dmm =(132 \pm 74) \times 10^{-6}$} respectively. 
The present analysis is based on \mbox{$\sim 10$ times} more lines, since the comprehensive fitting technique allows to fit partially overlapped and blended H$_{2}$ absorption features, and on more accurate rest wavelength measurements, which are considered exact for the purpose of comparison. This yielded a value of $\dmm$ which is \mbox{$\sim 10$ times} more precise than that previously found for this system \citep{Ivanchik2002}.

\begin{figure}
    \centering
    \includegraphics[width=\columnwidth]{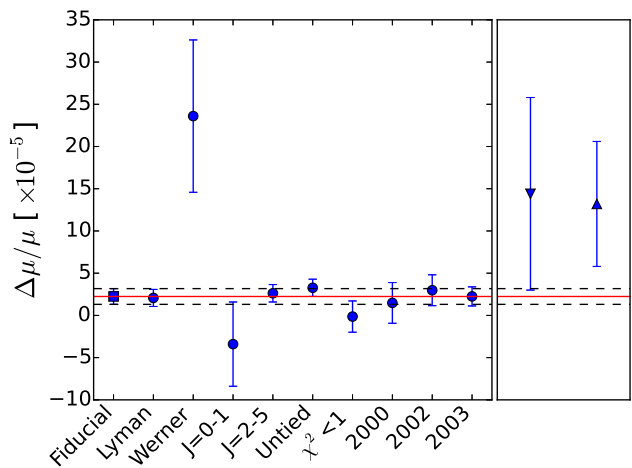}
    \caption{Left panel: Values of a varying $\mu$ obtained from the H$_{2}$ and HD absorption model, shown by a (blue) square, as well as from the consistency tests, shown by (blue) circles. The (red) solid line represents the fiducial $\dmm$ value and the two dashed lines show its $\pm 1\sigma$ boundaries. Right panel: The (blue) upwards and downwards triangles represent the two measurements derived by \protect\cite{Ivanchik2002} using the laboratory wavelengths reported by \protect\cite{Morton1976} and \protect\cite{Abgrall1993}, respectively.}
    \label{fig:test}
\end{figure}

Following the same approach of \cite{Bagdonaite2014}, a number of tests were performed to investigate the statistical robustness of the $\dmm$ value, hereafter the fiducial value. In the following four sections, four different tests are presented along with their results, which are shown in \mbox{Fig \ref{fig:test}}. Note that, in each test, the transitions that were not considered in the subset were included in the fit, but their $K_{i}$ coefficients were set to zero, i.e. they contributed to the absorption profile but not directly to a constraint on the $\dmm$ parameter.

\subsubsection{Isolating Lyman and Werner transitions}
\label{subsubsec:lyman}
Previous studies showed that, since the H$_{2}$ sensitivity coefficients are wavelength-dependent, the presence of a distortion in the UVES wavelength scale would be nearly degenerate with a variation in $\mu$ \citep{Ivanchik2005,Ubachs2007,Malec2010}. This degeneracy will, in principle, be broken to some extent by fitting together the Lyman and the Werner transitions, since they have different sensitivity coefficients at similar rest wavelengths. The effect of such wavelength-dependent distortions can be investigated by separating the two H$_{2}$ band systems. 

The two values returned are \mbox{$\dmm = (21 \pm 10_{stat}) \times 10^{-6}$} for the Lyman band systems and \mbox{$\dmm = (236 \pm 90_{stat}) \times 10^{-6}$} for the Werner band systems. The fact that the two values do not match within $2 \sigma$ is considered as a hint of the presence of systematic effects. The shift between the Lyman-only and the Werner-only $\dmm$ values requires a distortion slope which is \mbox{$\sim 10$ times} larger than what commonly found in UVES, hence it cannot be ascribed only to long-range distortions. This is considered evidence of an extra, unknown systematic effect which is affecting the fiducial value of $\dmm$. The larger uncertainty on the value of $\dmm$ delivered by the Werner transitions is ascribed to three effects: (i) the Werner transitions fall in the bluest part of the spectrum, which has a low S/N (\mbox{$\sim 10$} per 2.5 \kms\ per pixel at \mbox{3320 \AA}); (ii) the Werner transitions represent \mbox{$\sim 16\%$} of the total number of H$_{2}$ transitions considered; (iii) the spread in the sensitivity coefficients of the Werner transitions, $\Delta K_{W} = 0.02$, is lower than for Lyman transitions, $\Delta K_{L} = 0.05$.

\subsubsection{Isolating low and high J transitions}
\label{subsubsec:hot}
The impact of possible temperature inhomogeneities in the absorbing system on the constraint on $\dmm$ was estimated by fitting the cold, low-\emph{J} levels ($J < 2$) and the hot, high-\emph{J} levels ($J = 2 - 5$) separately. This is possible because, due to the para-ortho distribution, the $J = 1$ rotational state is significantly populated even at low temperatures, as discussed by \cite{Ubachs2007}. 

The two values derived from the cold and the warm transitions are \mbox{$\dmm = (-34 \pm 50_{stat}) \times 10^{-6}$} and \mbox{$\dmm = (26 \pm 10_{stat}) \times 10^{-6}$}, respectively. The larger error on the measurement derived from the colder states can be explained from the fact that there are only 15 transitions with $J = 0 - 1$ in the sample and they are nearly saturated, resulting in larger uncertainties on the position of the centroids of such absorption features. The two values match within their uncertainties, showing that the temperature inhomogeneities in the absorbing system do not significantly affect the constraint.

\subsubsection{Untying free parameters among different J-levels}
\label{subsubsec:untied}
As discussed in \mbox{Section \ref{sec:method}}, the present analysis was performed under the assumption that all the H$_{2}$ transitions share the same value of \emph{z} and \emph{b}. However, this may not be true if there is a spatial inhomogeneity in the absorbing cloud. In this case, transitions with different \emph{J}-levels will arise under different physical conditions, which would be reflected by different values of their Doppler width \emph{b}. 

To test the validity of this assumption, a fit was run with the free parameters \emph{z} and \emph{b} untied among the different \emph{J}-levels. The test returned \mbox{$\dmm = (5 \pm 10_{stat}) \times 10^{-6}$}, which is in a good agreement with the fiducial value. As a consequence, the assumption of spatial homogeneity used in this analysis has almost no impact on the constraint on $\dmm$.

\subsubsection{Excluding problematic regions}
\label{subsubsec:problematic}
Half of the selected spectral regions contain H$_{2}$ absorption features that show self-blending and overlap with intervening H~\textsc{i} or narrow, unidentified absorption features that are likely to be metal absorption features. Such intervening lines were included in the fit, but they do not benefit from the strength of the comprehensive fitting method, since they cannot be tied in any way to other transitions. Hence, any flaw in modelling these lines will affect the fiducial value of $\dmm$.

To test the impact of the unidentified intervening lines, a fit was performed considering only the 21 spectral regions that have a $\chi^{2}_{\nu} < 1$. The fit returned a value of \mbox{$\dmm = (-1 \pm 18) \times 10^{-6}$}, which matches with the fiducial value within their uncertainties. It is concluded that the H$_{2}$ absorption model is robust enough not to be affected by flaws in the model of intervening neutral hydrogen and metal transitions.

\subsubsection{Separating exposures from different periods}
\label{subsubsec:years}
UVES is known to suffer from wavelength calibration distortions that are likely to originate in the internal settings of the instrument, as discussed in \mbox{Section \ref{subsec:longrange}}. Such distortions are likely to affect the fiducial value of $\dmm$, hence quantifying this effect is crucial for any $\mu$ variation analysis. The attempt made to distortion-correct the spectrum is presented in \mbox{Section \ref{subsec:longrange}}, while here a test to estimate the consistency of the distortion slope across the observational programs is presented. 

The final Q1232+082 spectrum was divided into 3 `subspectra' -- combined spectra formed from subsets of exposures -- each comprising observations taken only in 2000, 2002, and 2003. Any change in the magnitude of the wavelength distortions would be reflected in a significant variation of the $\dmm$ values derived from the subspectra. A value for $\dmm$ was derived from each subspectrum, returning $\dmm = (15 \pm 24_{stat}) \times 10^{-6}$ for exposures taken in 2000, $\dmm = (30 \pm 18_{stat}) \times 10^{-6}$ for 2002, and $\dmm = (22 \pm 11_{stat}) \times 10^{-6}$ for exposures taken in 2003. The differences in the errors among the values reflect the fact that the subsets do not have the same integration time, hence the S/N determining the statistical error changes among the subsets. The measurements agree well within their uncertainties, showing that the relative differences in the telescope settings among the different exposures do not significantly affect the fiducial value of $\dmm$.

\section{Systematic error}
\label{sec:systematic}
Following from \mbox{Eq. (\ref{eq:shift})}, the measurable effect of a non-zero $\dmm$ is a shift between molecular absorption features. As a consequence, any effect that introduces a distortion of the wavelength scale is likely to introduce a systematic error on the measured $\dmm$ value. Note that this holds only for wavelength-dependent distortions, since any constant velocity offset will not cause any relative shift between the absorption features. Such a systematic will affect the resulting value for the absorption redshift parameter. The role of wavelength calibration errors as well as the spectral redispersion on the fiducial value of $\dmm$ was studied in previous works on different systems \citep{Malec2010,King2011,Bagdonaite2012,Bagdonaite2014,Dapra2015}, while lack of attached ThAr calibrations was studied by \cite{Bagdonaite2012,Bagdonaite2014}. In the following, the contributions to the systematic error budget from four sources of systematics are discussed following the same approach that was used in the aforementioned works. 

\subsection{Long-range distortions}
\label{subsec:longrange}
The UVES spectrograph is known to suffer from wavelength calibration distortions. Long-range distortions were first detected by \cite{Rahmani2013} and, more recently, analyzed in further detail by \cite{Whitmore2015}. Such distortions are likely to be due to a difference in the path the light from the quasar and from the calibration lamp travels within the instrument. 

The supercalibration technique \citep{Whitmore2015} is a powerful method to correct the spectrum for the long-range distortions and was successfully applied to other systems \citep{Bagdonaite2014,Dapra2015,Dapra2016}. It consists in a comparison of a spectrum taken with UVES with a reference spectrum from a Fourier Transform Spectrometer (FTS) solar spectrum, which has a more accurate frequency scale \citep{Chance2010}\footnote{Available at \url{http://kurucz.harvard.edu/Sun/irradiance2005/irradthu.dat}}. Typical supercalibration targets are asteroids and `solar twin' stars. The former have the same spectrum as the Sun since they reflect the solar light, while the latter are objects that show a spectrum which is almost identical to the solar one \citep{Melendez2009,Datson2014}.

Since no supercalibration targets could be found in the ESO archive in a temporal window of \mbox{$\sim 2$ weeks} of the quasar exposures, the technique is not fully applicable to Q1232+082. However, the sparse sample reported by \cite{Whitmore2015} suggests that the UVES wavelength distortions might be quasi-stable in the period 2004-2008, with a distortion slope of \mbox{$\sim 200$ \ms}\ per \mbox{1000 \AA}. Moreover, \cite{Whitmore2015} found one supercalibration target in 2001 delivering a distortion slope of \mbox{$\sim 100$ \ms}\ per \mbox{1000 \AA}. Assuming that the distortions affecting UVES were stable in the period 2000-2003, the slope value of \mbox{$\sim 100$ \ms} per \mbox{1000 \AA} was used to distortion-correct the Q1232+082 spectrum, delivering an updated fiducial $\dmm$ value of \mbox{$\dmm = (19 \pm 9_{stat}) \times 10^{-6}$}. An uncertainty on the distortion correction of \mbox{$\pm 100$ \ms} per \mbox{1000 \AA} was estimated from the spread between the distortion slope values in 2001 and in the period 2004-2008. This translates into a systematic uncertainty on $\dmm$ of \mbox{$\sim 5 \times 10^{-6}$ ($1 \sigma$)} and it was added to the systematic error budget. The presence of the long-range distortions with a positive distortion slope is found to push the $\dmm$ value towards a more positive value, as was found in other systems \citep{Rahmani2013,Bagdonaite2014,Bagdonaite2015,Dapra2015}.

\subsection{Intra-order distortions}
\label{subsec:intraorder}
At small scales, within single echelle orders, wavelength calibration distortions up to \mbox{$\sim 100$ \ms} are expected in the spectra taken with UVES \citep{Griest2010,Whitmore2010}. However, since the H$_{2}$ transitions are spread over multiple echelle orders, such distortions are not expected to be dominant in the systematic error budget.

To estimate the impact of such intra-order distortions on $\dmm$, a sawtooth wavelength distortion with an amplitude of \mbox{$\pm 100$ \ms} was introduced in each echelle order before combining them together into the final quasar spectrum. A fit of the artificially distorted spectrum returned a value of \mbox{$\dmm = (22 \pm 9) \times 10^{-6}$}, which is well in agreement with the fiducial value before correcting for the long-range distortions. As a result, the intra-order distortions' contribution to the systematic uncertainty is \mbox{$< 1 \times 10^{-6}$}, which is of the same order of that found in previous, similar studies on different systems \citep{Malec2010,King2011,Weerdenburg2011,Bagdonaite2014,Dapra2015}. This effect, negligible compared to the uncertainty introduced by the long-range distortions, was included in the systematic error budget.

\subsection{Lack of attached ThAr calibrations}
\label{subsec:thar}
The Q1232+082 spectra available in the ESO archive were recorded without the use of the attached ThAr calibration lamp exposure, hence they were calibrated using the standard ThAr calibration taken at the end of the night. Previous studies on different systems reported that the lack of attached ThAr calibrations introduces an error on the final value of $\dmm$ which \mbox{is $< 1 \times 10^{-6}$} \citep{Bagdonaite2014,Dapra2015}, which was added to the systematic error budget. 

\subsection{Spectral redispersion}
\label{subsec:redispersion}
The final Q1232+082 spectrum is composed by adding several exposures together. This procedure implies a redispersion of all the individual exposures on a common wavelength grid. The rebinning can cause flux correlations between neighbouring pixels, while the choice of the grid can distort the line-profile shapes affecting the value of $\dmm$.

The impact of the spectral redispersion on $\dmm$ was investigated by deriving a value of $\dmm$ from six Q1232+082 spectra constructed using six different velocity grids in the range \mbox{$2.47 - 2.53$ \kms} per pixel. The average deviation from the fiducial value is \mbox{$1.4 \times 10^{-6}$} and was included in the systematic uncertainty budget.

\subsection{Total systematic uncertainty}
\label{subsec:totalsyst}
The total systematic uncertainty on the fiducial value was derived by adding in quadrature all the contributions to the systematic error budget. The returned systematic error is $\sim 5 \times 10^{-6}$, therefore the updated fiducial value becomes \mbox{$\dmm = (19 \pm 9_{stat} \pm 5_{syst}) \times 10^{-6}$}.

\section{Conclusions}
\label{sec:conclusions}
A re-analysis of molecular hydrogen absorption at \mbox{$\zab \simeq 2.34$} towards quasar Q1232+082 is presented, in order to constrain a possible temporal variation of $\mu$ over cosmological time scales. The absorption system shows 106 H$_{2}$ and HD transitions associated with the DLA feature and spread over a range of \mbox{$\sim 400$ \AA}. The dataset contains strongly saturated absorption features as well as overlaps with intervening lines. The comprehensive fitting technique used is able to handle such overlaps, as well as the partial coverage of the absorption system, whose respective effects are included in the absorption model. The H$_{2}$ and HD absorption was used to constrain a possible $\mu$-variation. 

Since no supercalibration exposures are available for the Q1232+082 spectrum, an attempt to correct the spectrum for the long-range distortions was made based on the very sparely-sampled information from \cite{Whitmore2015}. An estimate of the uncertainty on the distortion correction was made, based on the same sample, and it was added to the systematic error budget. The contributions to the systematic error budget from four different sources were investigated and the total systematic uncertainty was found to be dominated by the long-range wavelength distortions. The fiducial $\dmm$ value delivered by the analysis is \mbox{$\dmm = (19 \pm 9_{stat} \pm 5_{syst}) \times 10^{-6}$}.

This value is \mbox{$\sim 10$ times} more precise than those reported by \cite{Ivanchik2002}. This is due to several improvements: (i) the comprehensive fitting technique applied here allows to expand the transition sample by including partially overlapped absorption features, (ii) the spectrum of Q1232+082 was built including two observational programs from 2003, that contribute additional \mbox{$\sim 8$ hrs} of integration, (iii) \cite{Ivanchik2002} used two sets of rest wavelengths with a low fractional accuracy, which is reflected in the uncertainty on the derived values of $\dmm$. The updated fiducial value of $\dmm$ presented in this work was derived using accurate rest wavelengths from laser spectroscopy \citep{Hollenstein2006,Ivanov2008,Salumbides2008}. Given the high accuracy of the studies, the rest wavelengths can be considered exact for the purpose of the comparison, (iv) the updated fiducial value of $\dmm$ was derived after including a correction for the partial coverage in the H$_{2}$ absorption model.

\begin{figure}
    \centering
    \includegraphics[width=\columnwidth]{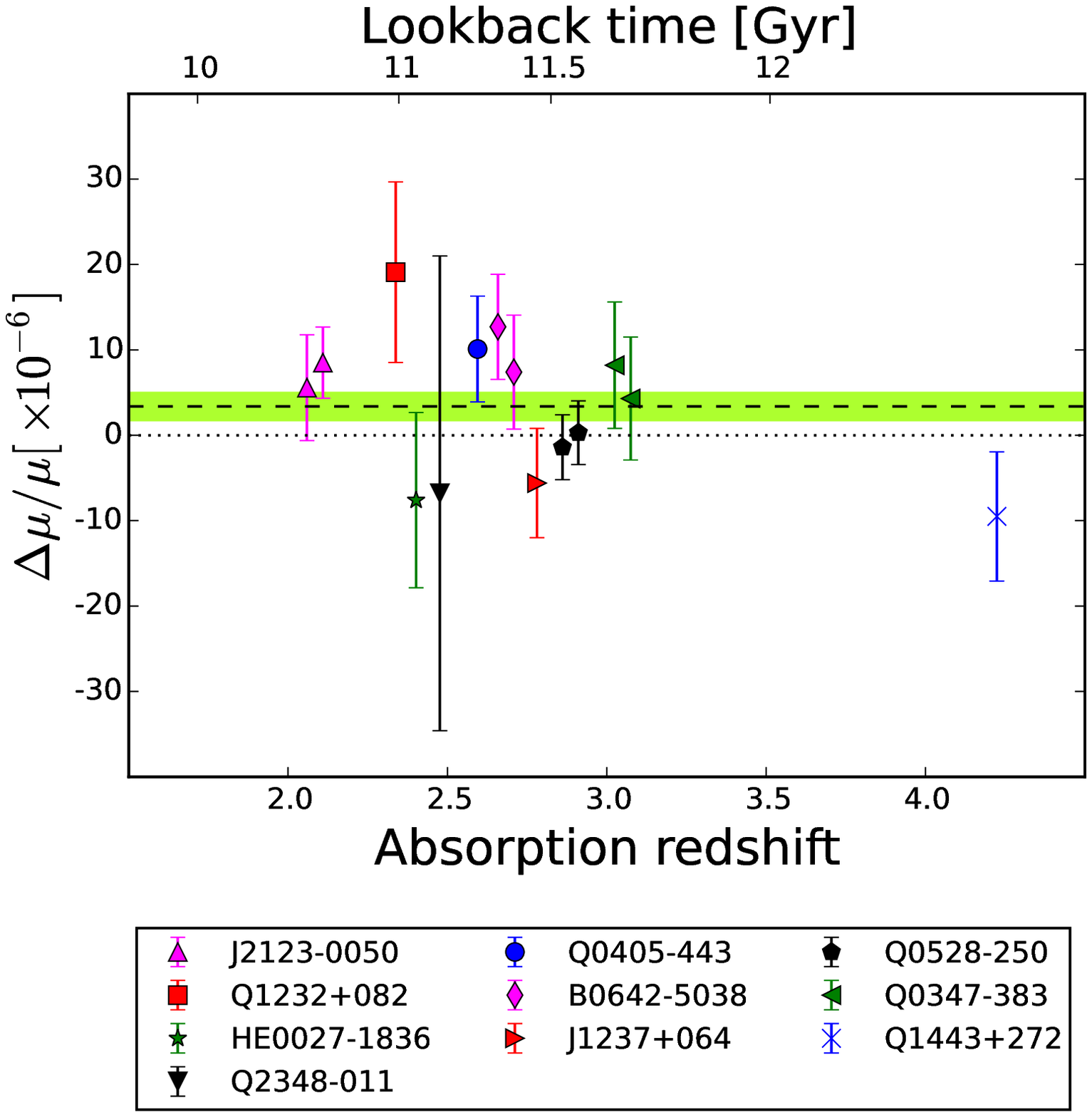}
    \caption{Measurements of $\dmm$ derived from molecular hydrogen absorbing systems. The value presented here is shown with a (red) square. The other values of $\dmm$ were derived from systems: J2123-0050 \citep{Malec2010,Weerdenburg2011}, HE0027-1836 \citep{Rahmani2013}, Q2348-011 \citep{Bagdonaite2012}, Q0405-443 \citep{King2008}, B0642-5038 \citep{Albornoz2014,Bagdonaite2014}, J1237+0647 \citep{Dapra2015,Dapra2016}, Q0528-250 \citep{King2008,King2011}, Q0347-383 \citep{King2008,Wendt2012} and Q1443+272 \citep{Bagdonaite2015}. Note that multiple $\dmm$ measurements were derived from systems J2123-0050, B0642-5038, Q0528-250 and Q0347-383, and are presented with an offset of $+0.05$ on the x axis to avoid overlaps. The dashed line represents the weighted mean of all the values, the shaded area shows its $\pm 1 \sigma$ boundaries, and the dotted line represents the zero level.
    }
    \label{fig:overview}
\end{figure}

The result presented here can be compared with other studies on different H$_{2}$ absorption systems, as shown in \mbox{Fig. \ref{fig:overview}}. The weighted mean of all the $\dmm$ values considered results in a value of \mbox{$\dmm = (3.4 \pm 1.6) \times 10^{-6}$}, which is consistent with no variation over a look-back time of \mbox{$\sim 10.5$-12.5 Gyrs} at the $3 \sigma$ level. It is noted that not all the measurements were corrected for the effect of wavelength distortions, which, for most cases, pushes the measured $\dmm$ value towards more positive values by few parts per million. As a consequence, the averaged constraint should be considered as an upper limit to the temporal variation of $\mu$ over cosmological timescales.

A list of the known systems showing H$_{2}$ and/or CO absorption was compiled by \cite{Ubachs2016}. The ten best absorbers for a $\mu$ variation analysis, because of their brightness, with a Bessel $R_{mag} \le 18.4$, and H$_{2}$ column density, $\log N_{H_{2}} \ge 14.5$, are shown in \mbox{Fig. \ref{fig:overview}}. With the analysis presented here, we complete a set of the ten best absorbers for $\mu$ variation. Note that the set does not contain the system  Q2100--0641 at $\zab = 3.09$, which has $R_{mag} = 17.52$ and $\log N_{H_{2}} = 18.76$ \citep{Balashev2015}, and could represent a good target for future investigations. 

The current constraint on $\dmm$ is almost equally limited by statistical and systematic uncertainties. While the former can be improved by increasing the S/N, the latter is dominated by the long-range wavelength distortions. Such distortions affects not only UVES, but also the High Resolution Echelle Spectrograph, HIRES, mounted on the Keck telescope, which is the other instrument used to investigate a temporal drift of $\mu$. \cite{Whitmore2015} expanded and refined the supercalibration technique to distortion-correct the quasar spectra, and they showed that the long-range distortions are ubiquitous across the entire UVES and HIRES history. The fact that most of the archival exposures do not have any attached supercalibration precludes a significant reduction in the systematic error on $\dmm$ when using such data from these spectrographs. 

The current constraint on a varying $\mu$ can be improved with the new generation of  high resolution spectrographs, like the upcoming Echelle SPectrograph for Rocky Exoplanet and Stable Spectroscopic Observations \citep[ESPRESSO,][]{Pepe2010}. ESPRESSO, that has a resolution $\sim 3$ times higher than the one used in this work, and is able to resolve VCs with a Doppler broadening parameter as narrow as \mbox{$\sim 1.0$ \kms}. Moreover, being immune, in principle, to the long-range wavelength distortions, thanks to its fibre feed and frequency-comb wavelength calibration, it will deliver constraints on $\dmm$ with smaller systematic uncertainties. However, ESPRESSO will not cover wavelengths shorter than \mbox{$\sim 3800$ \AA}, which means it can only target H$_{2}$ absorption lines in systems at \mbox{$\zab > 2.4$}, and very few such systems are known towards bright quasars \citep{Ubachs2016}.

\section*{Acknowledgments}
The authors thank the Netherlands Foundation for Fundamental Research of Matter (FOM) for financial support. MTM thanks the Australian Research Council for \textsl{Discovery Project} grant DP110100866 which supported this work. The work is based on observations with the ESO Very Large Telescope at Paranal (Chile). WU thanks the European Research Council for an ERC-Advanced grant (No 670168).

\appendix
\section{Absorption model}
\label{app:model}
\begin{figure*}
    \centering
    \includegraphics[width=2\columnwidth]{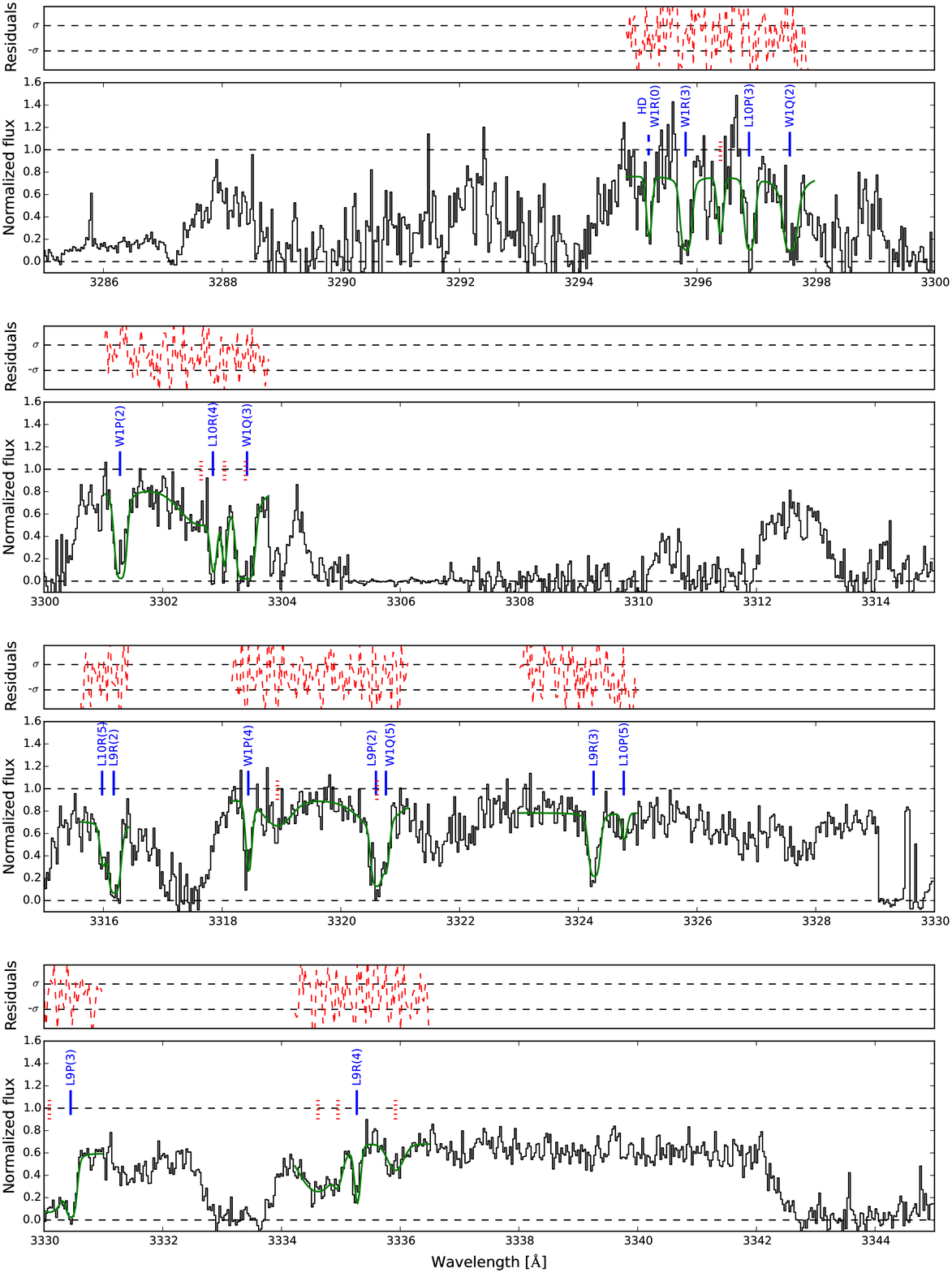}
    \caption{Spectrum of quasar Q1232+082 (part 1 of 8). The H$_{2}$ absorption model is represented with a (green) solid line. H$_{2}$ transitions are shown with (blue) solid ticks, HD transitions with (blue) dashed ticks, and the (red) dotted ticks show the position of the intervening H~\textsc{I} and metal lines.}
    \label{fig:spectrum1}
\end{figure*}

Figures \ref{fig:spectrum1}-\ref{fig:spectrum6} show the H$_{2}$ absorption model in the spectrum of the absorbing system at $\zab \simeq 2.34$ towards quasar Q1232+082.

\begin{figure*}
    \centering
    \includegraphics[width=2\columnwidth]{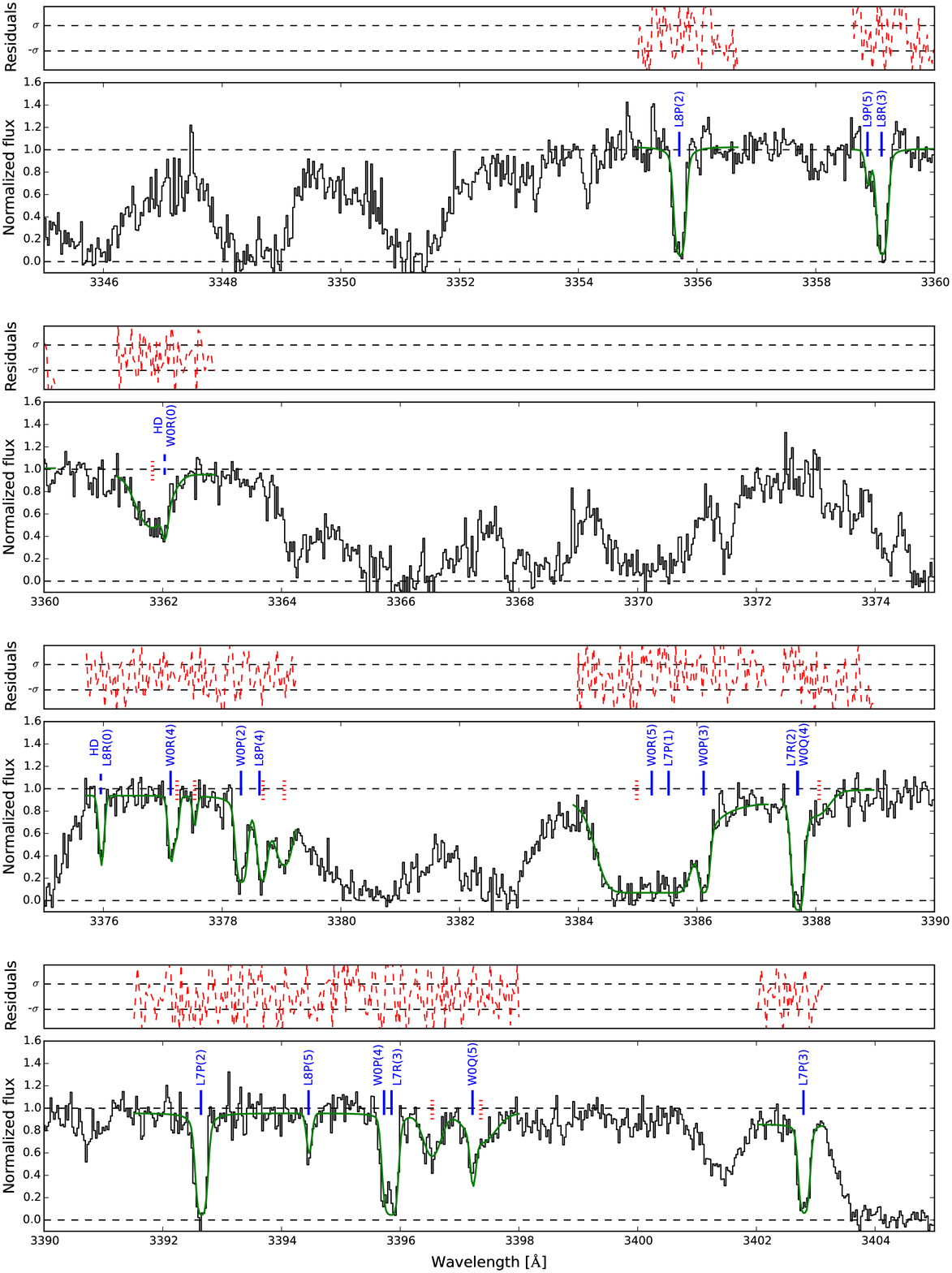}
    \caption{Spectrum of quasar Q1232+082 (part 2 of 8), continued.}
    \label{fig:spectrum2}
\end{figure*}

\begin{figure*}
    \centering
    \includegraphics[width=2\columnwidth]{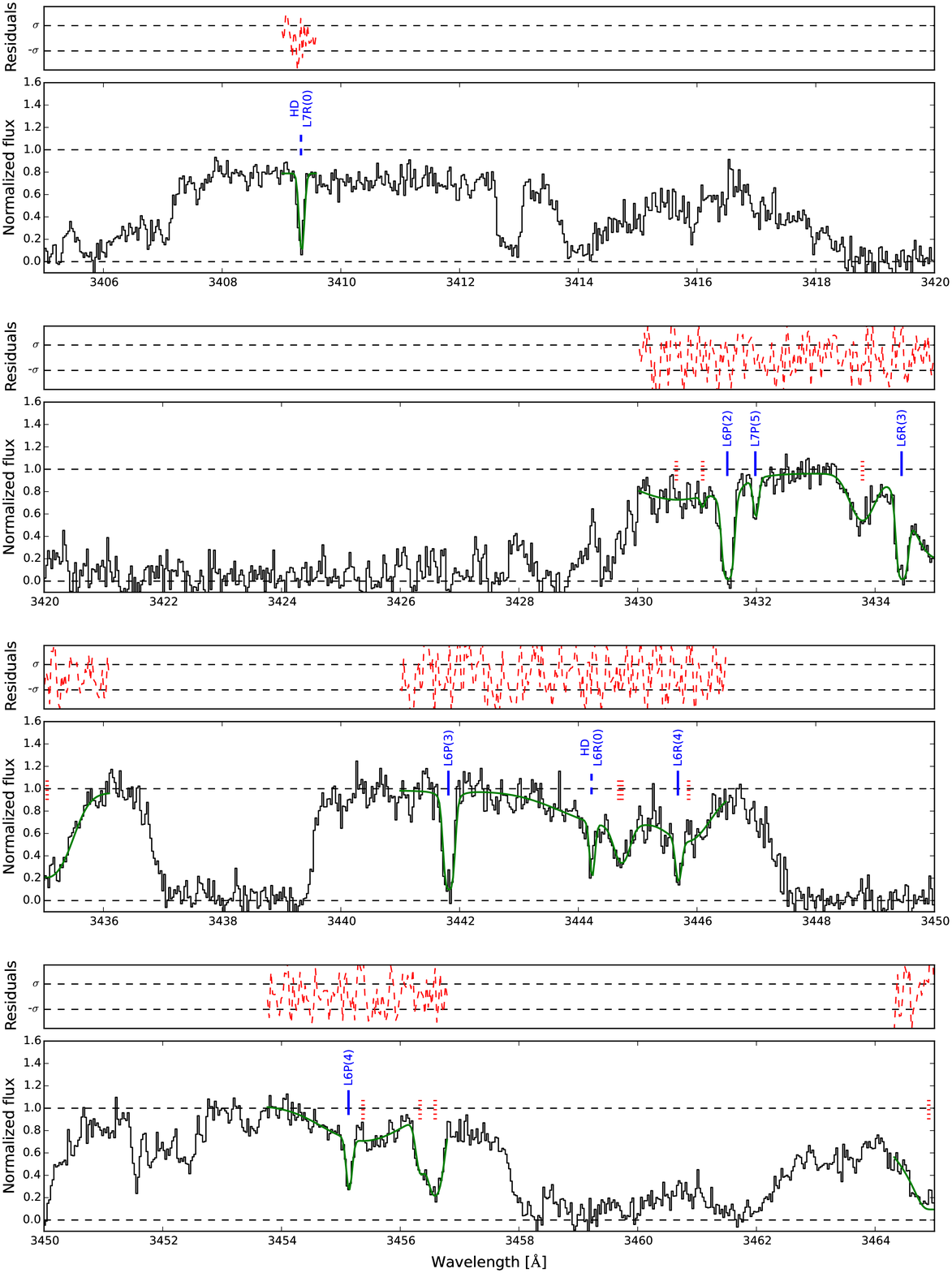}
    \caption{Spectrum of quasar Q1232+082 (part 3 of 8), continued.}
    \label{fig:spectrum3}
\end{figure*}

\begin{figure*}
    \centering
    \includegraphics[width=2\columnwidth]{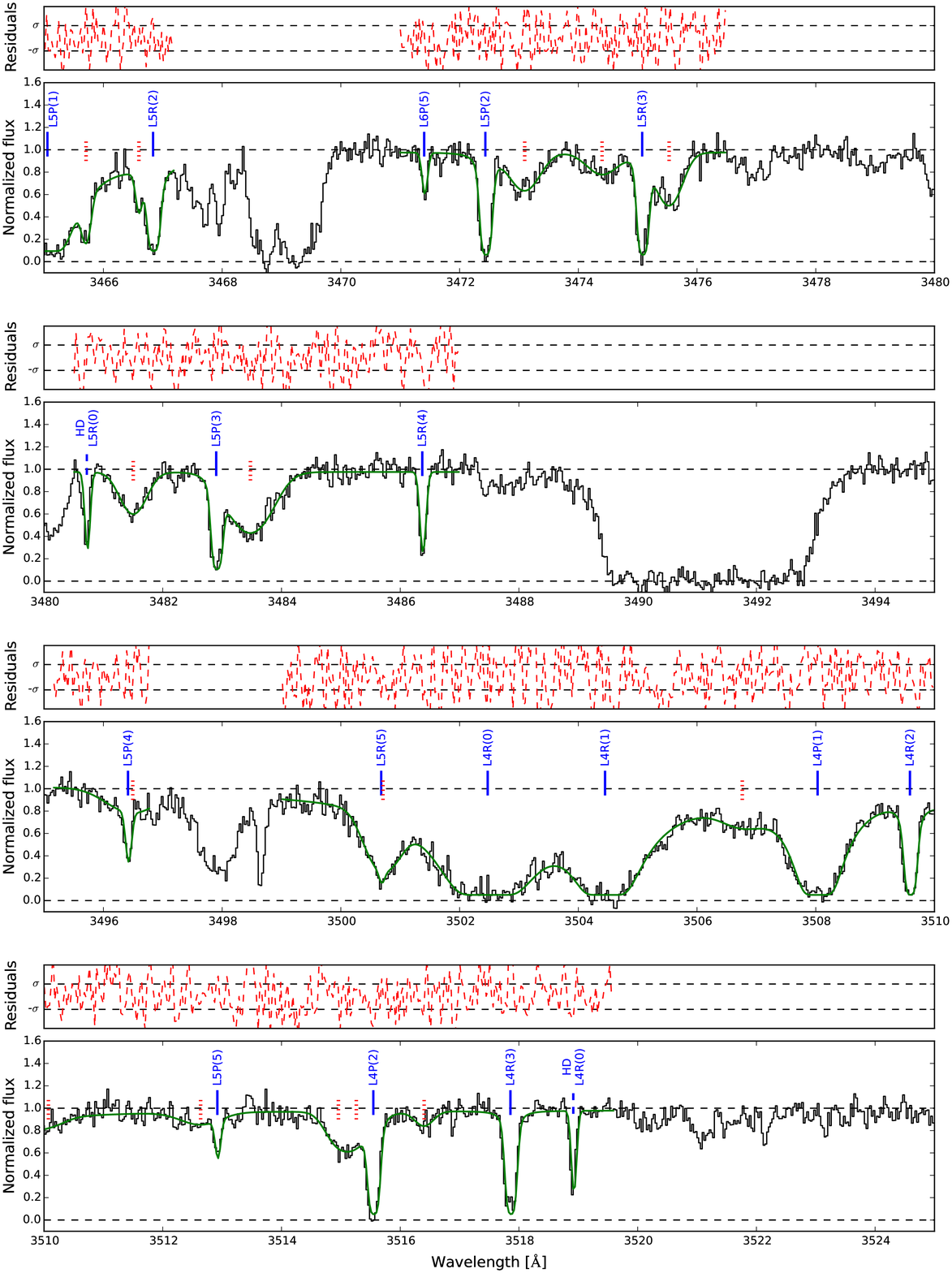}
    \caption{Spectrum of quasar Q1232+082 (part 4 of 8), continued.}
    \label{fig:spectrum4}
\end{figure*}

\begin{figure*}
    \centering
    \includegraphics[width=2\columnwidth]{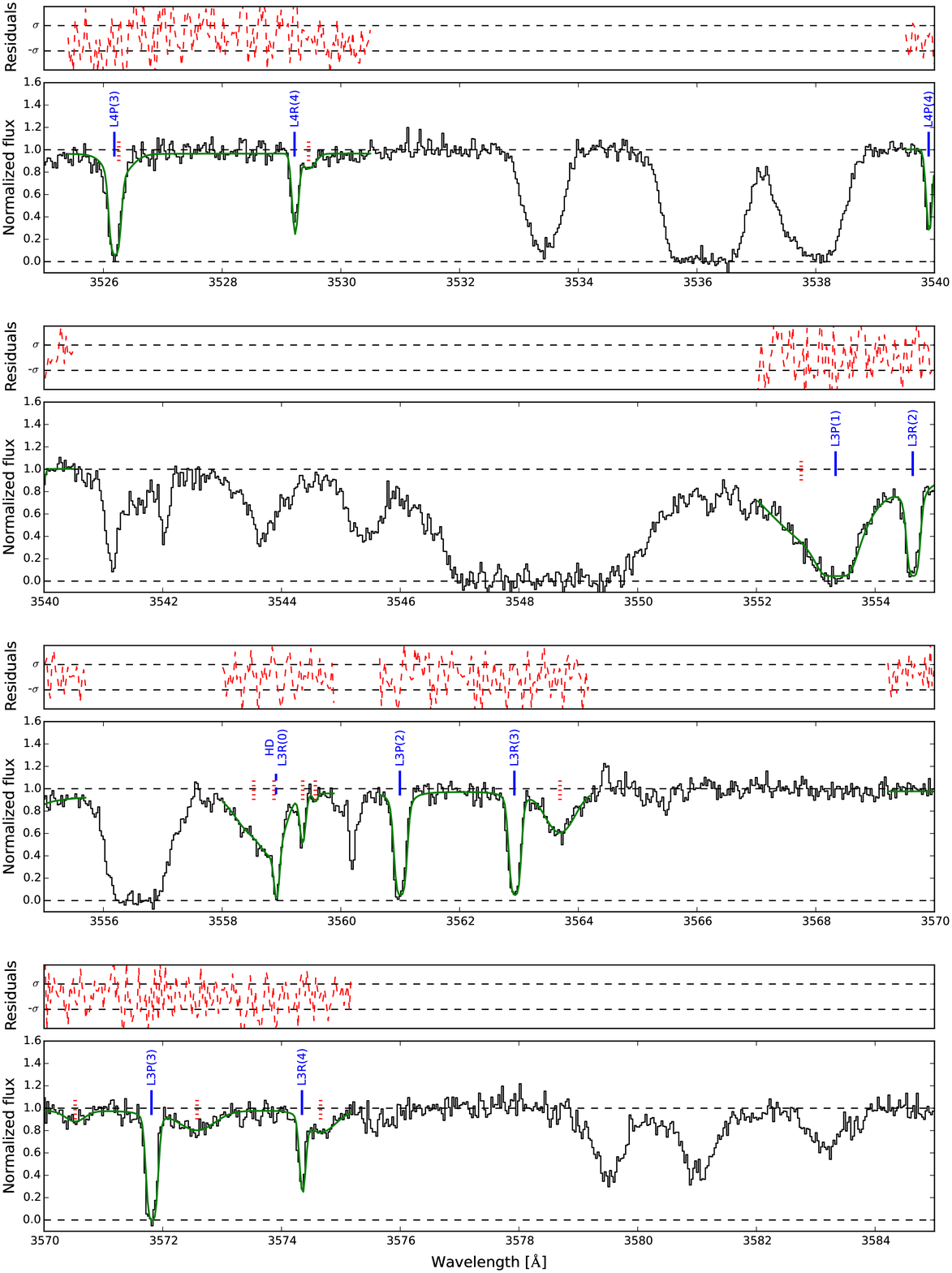}
    \caption{Spectrum of quasar Q1232+082 (part 5 of 8), continued.}
    \label{fig:spectrum5}
\end{figure*}

\begin{figure*}
    \centering
    \includegraphics[width=2\columnwidth]{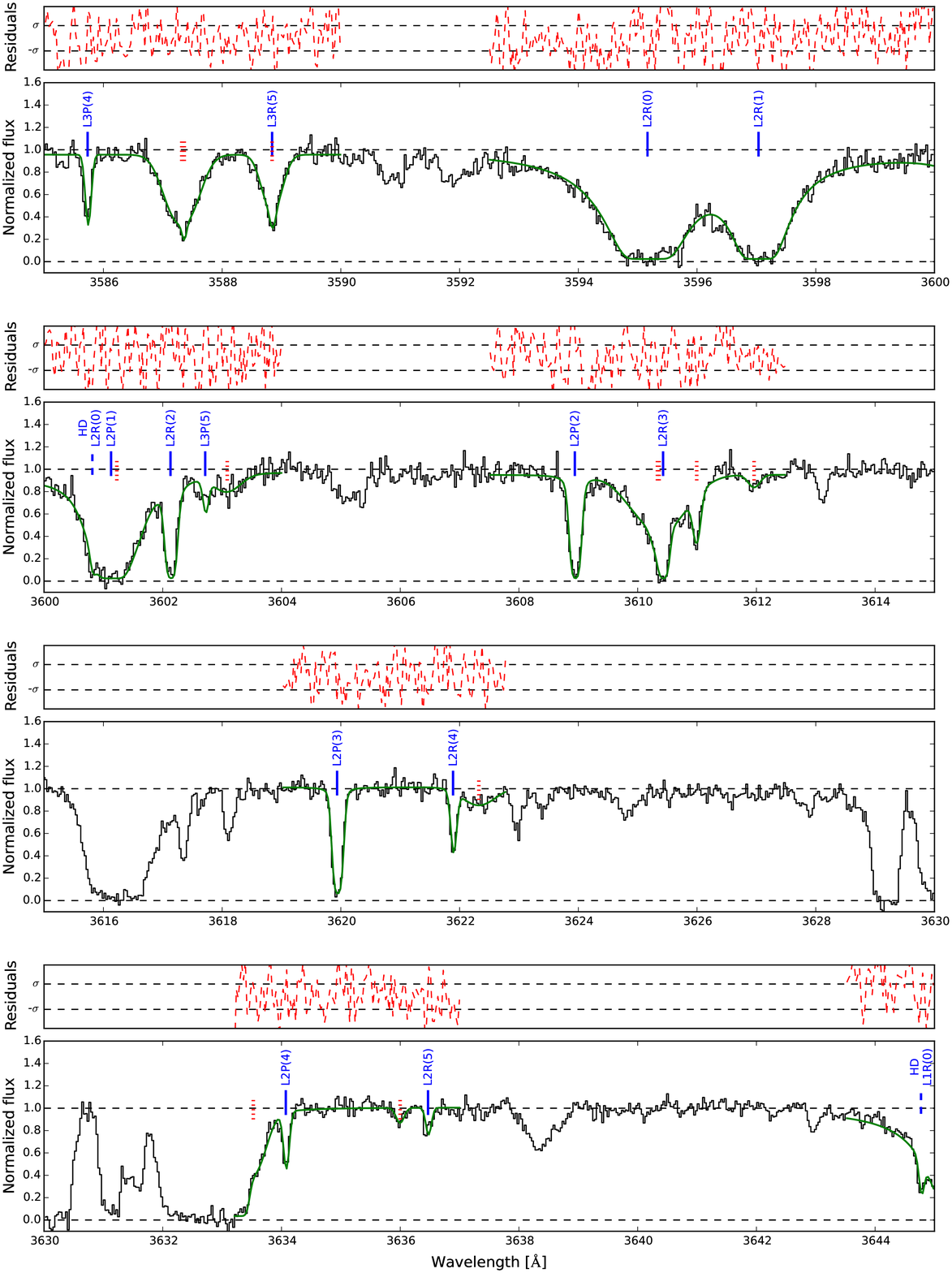}
    \caption{Spectrum of quasar Q1232+082 (part 6 of 8), continued.}
    \label{fig:spectrum6}
\end{figure*}

\begin{figure*}
    \centering
    \includegraphics[width=2\columnwidth]{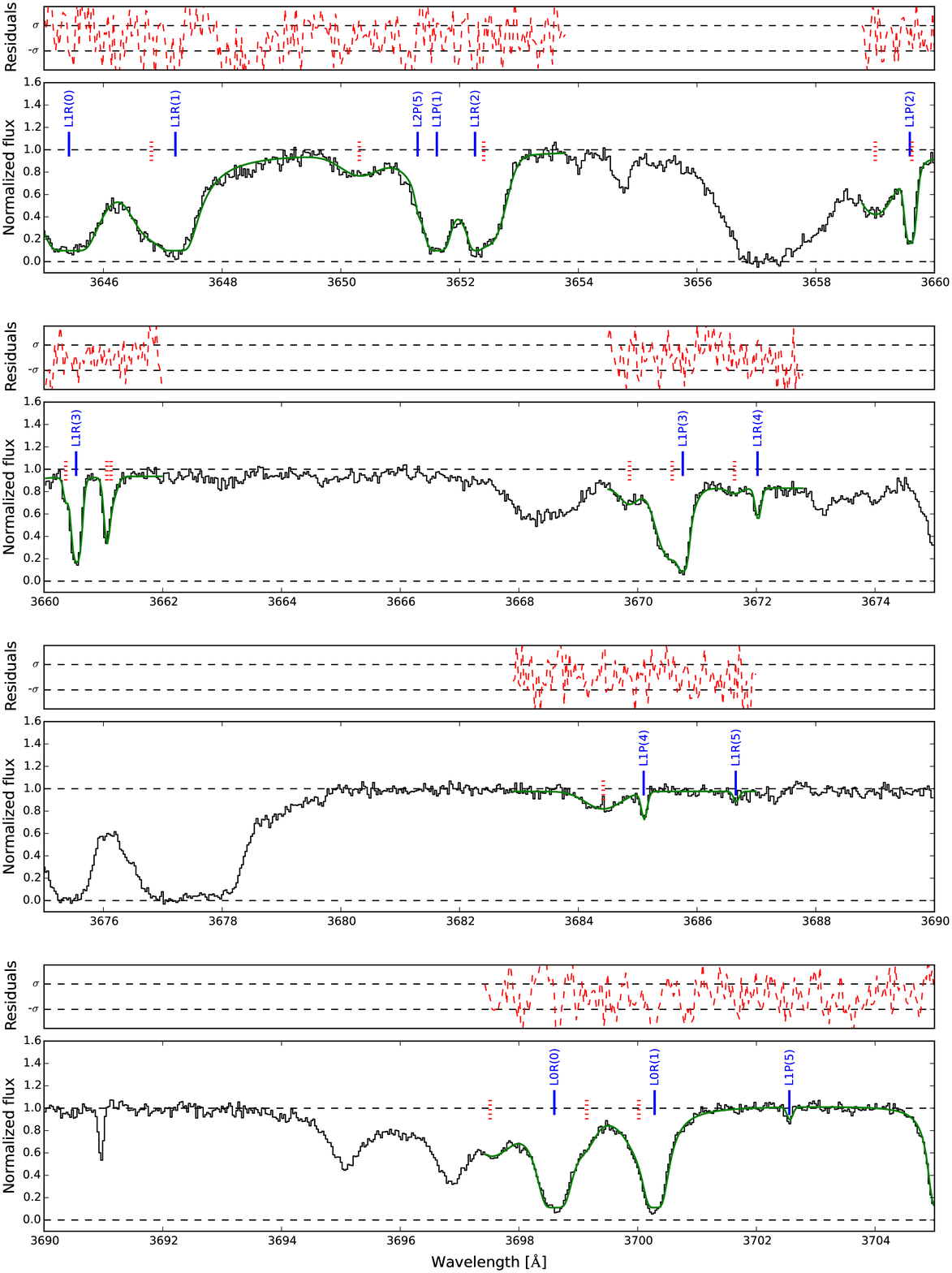}
    \caption{Spectrum of quasar Q1232+082 (part 7 of 8), continued.}
    \label{fig:spectrum6}
\end{figure*}

\begin{figure*}
    \centering
    \includegraphics[width=2\columnwidth]{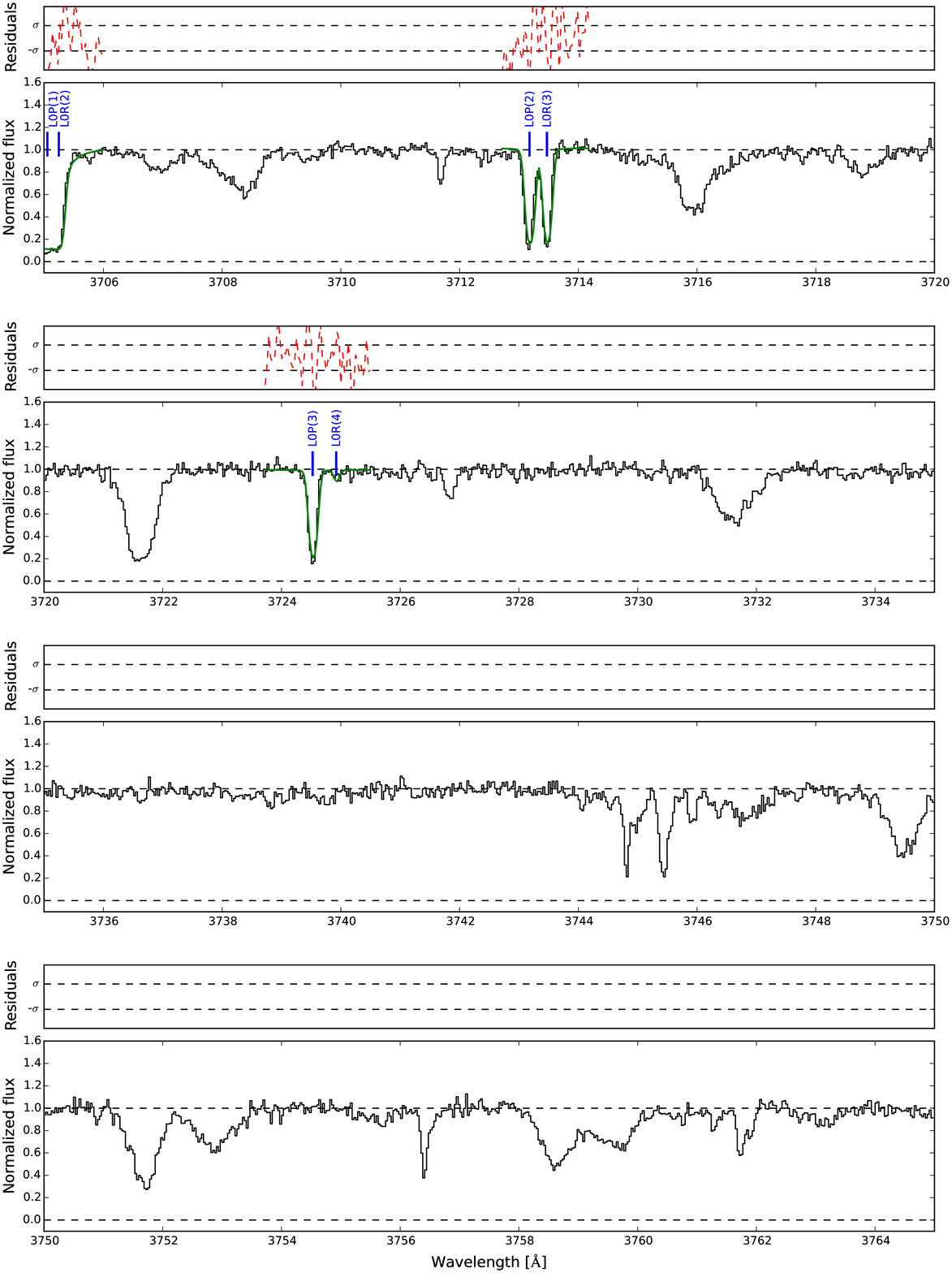}
    \caption{Spectrum of quasar Q1232+082 (part 8 of 8), continued.}
    \label{fig:spectrum6}
\end{figure*}

\bsp
\label{lastpage}

\end{document}